\documentstyle[mprocl,psfig]{article}

\begin{document}

\arraycolsep1.5pt 

\def\Journal#1#2#3#4{{#1} {\bf #2}, #3 (#4)}

\def\NCA{\em Nuovo Cimento}
\def\NIM{\em Nucl. Instrum. Methods}
\def\NIMA{{\em Nucl. Instrum. Methods} A}
\def\NPB{{\em Nucl. Phys.} B}
\def\PLB{{\em Phys. Lett.}  B}
\def\PRL{\em Phys. Rev. Lett.}
\def\PRD{{\em Phys. Rev.} D}
\def\ZPC{{\em Z. Phys.} C}
\def\st{\scriptstyle}
\def\sst{\scriptscriptstyle}
\def\mco{\multicolumn}
\def\epp{\epsilon^{\prime}}
\def\vep{\varepsilon}
\def\ra{\rightarrow}
\def\ppg{\pi^+\pi^-\gamma}
\def\vp{{\bf p}}
\def\ko{K^0}
\def\kb{\bar{K^0}}
\def\al{\alpha}
\def\ab{\bar{\alpha}}
\def\be{\begin{equation}}
\def\ee{\end{equation}}
\def\bea{\begin{eqnarray}}
\def\eea{\end{eqnarray}}
\def\CPbar{\hbox{{\rm CP}\hskip-1.80em{/}}}

\def\sgreat{\lower2pt\hbox{$\buildrel {\scriptstyle >}
   \over {\scriptstyle\sim}$}}

%
%


\title{X-RAY NOVAE AND THE EVIDENCE FOR BLACK HOLE EVENT HORIZONS}

\author{RAMESH NARAYAN, MICHAEL R. GARCIA, JEFFREY E. MCCLINTOCK}

\address{Harvard-Smithsonian Center for Astrophysics\\60 Garden
Street, Cambridge, MA 02138, U.S.A.\\Email: rnarayan@cfa.harvard.edu, 
mgarcia@..., jmcclintock@...}


\maketitle\abstracts{We discuss new observations of X-ray novae which
provide strong evidence that black holes have event horizons.  Optical
observations of 13 X-ray novae indicate that these binary stars
contain collapsed objects too heavy to be stable neutron stars.  The
objects have been identified as black hole candidates.  X-ray
observations of several of these X-ray novae in quiescence with the
{\it Chandra} X-ray Observatory show that the systems are
approximately 100 times fainter than nearly identical X-ray novae
containing neutron stars.  The advection-dominated accretion flow
model provides a natural explanation for the difference.  In this
model, the accreting gas reaches the accretor at the center with a
large amount of thermal energy.  If the accretor is a black hole, the
thermal energy will disappear through the event horizon, and the
object will be very dim.  If the accretor is a neutron star or any
other object with a surface, the energy will be radiated from the
surface, and the object will be bright.  We discuss alternate
interpretations of the data that eliminate the need for
advection-dominated accretion.  Most of these alternatives still
require an event horizon to explain the unusually low X-ray
luminosities of the black hole candidates.  Some of the alternatives
are also inconsistent with observations.}

\section{Introduction}

One of the most exciting quests in high energy astrophysics --- indeed
in all of astrophysics --- is the search for black holes.  More than a
dozen excellent black hole candidates have been discovered in X-ray
binaries (\S2), and new candidates are being discovered at an
accelerating rate.  The candidate black holes are identified on the
basis of their masses which are found to be greater than the maximum
mass of a neutron star.

While the mass criterion is an excellent technique for finding black
hole candidates, one would like to have more direct evidence that the
candidates truly are black holes.  Ideally, one would like to be able
to show that a candidate has an {\it event horizon}, the defining
characteristic of a black hole.  We have pursued this goal for the
last several years.

Our approach has been to compare black hole X-ray binaries and neutron
star X-ray binaries under similar conditions and to show that there is
a large difference in the luminosity or some `other basic observational
property of the two systems, a difference that is most easily
explained by invoking an event horizon in the candidate black holes.
We describe in \S3 the observational data we have focused on and
discuss in \S4 our preferred explanation of the data, which requires
black hole candidates to have event horizons.  A number of groups have
come up with alternate explanations of the data.  We present these
ideas in \S5 and discuss the strengths and weaknesses of each
proposal.

\section{Candidate Black Holes and Neutron Stars in X-ray Binaries}

There are about 200 cataloged X-ray binaries in the Galaxy (van
Paradijs 1995), each containing either a neutron star or a black hole
accreting material from a companion star.  Most of these systems are
persistently bright in the X-ray band.  However, there are about 30
transient X-ray binaries which are known as ``X-ray novae'' or ``soft
X-ray transients'' (Tanaka \& Shibazaki 1996).  These systems are of
special interest in the search for black hole event horizons.

X-ray novae are characterized by episodic outbursts at X-ray, optical
and radio frequencies, which are separated by long intervals (years to
decades) of quiescence (Tanaka \& Shibazaki 1996; van Paradijs \&
McClintock 1995).  The outburst is caused by a sudden dramatic
increase in the rate of mass accretion onto the compact primary
(Lasota 2001).  The X-ray flux rises on a time scale of the order of
days, and subsequently declines on a time scale of weeks or months.
The X-ray flux in outburst can be several million times the quiescent
X-ray flux.

In the quiescent state, the absorption-line velocities of the
secondary star can be determined precisely because the non-stellar
light from the accretion flow is modest compared to the light of the
secondary.  These velocity data {\it vs.}~orbital phase determine the
value of the mass function, $f(M)$, which is an absolute lower limit
to the mass of the compact primary:

\begin{equation}
f(M) \equiv {M_{1}^{3}\sin^{3}i \over (M_{1}+M_{2})^{2}} = {PK^{3}
\over 2\pi G} < {M_{1}},
\end{equation}

\noindent where $M_{1}$ and $M_{2}$ are the masses of the compact
primary and the secondary star, respectively, K is the semiamplitude
of the observed velocity curve of the secondary, P is the orbital
period, $i$ is the orbital inclination angle, and G is Newton's
constant.

Reliable values of the mass function have been obtained for the 13
X-ray novae listed in Table 1.  For the first 11 systems, $M_{1} >
f(M)\ \sgreat\ 3M_{\odot}$.  Assuming that general relativity applies,
the maximum mass of a neutron star can be calculated to be about
$2-3M_\odot$ (Rhoades \& Ruffini 1974; Cook et al. 1994; Kalogera \&
Baym 1996), so we can be almost certain that the compact primaries in
these 11 systems are black holes.  The estimated values of $M_{1}$
given in the third column of Table 1 are based on additional
measurements of $i$ and $M_{2}$ and are significantly less certain
than the values of the mass function.

A number of X-ray novae also contain neutron star primaries, as
evidenced by the observation of type I X-ray bursts (Lewin, van
Paradijs, \& Taam 1995).  A type I burst, which is a firm signature of
a neutron star, is due to a thermonuclear flash in material that has
been freshly accreted onto the star's surface.  The phenomenology of
these bursts has been studied since their discovery in 1975. The burst
rise times are $\sim 1-10$~s and their decay times are $\sim 10$~s to
minutes.  The time-dependent spectra of X-ray bursts are well
described by a blackbody spectrum that cools as the burst intensity
decays.  Occasionally, intense flat-topped burst profiles are observed
for which the source luminosity reaches the Eddington limit (defined
in \S3).

In order to fairly compare neutron star X-ray novae (NSXN) and black
hole X-ray novae (BHXN), we must know their orbital periods, as this
has a large influence on the mass transfer rates (\S3).  This limits
our current sample of neutron star X-ray novae to SAX J1808.4--3658,
EXO 0748--676, 4U2129+47, MXB 1659--298, Cen X--4, Aql X--1, and
H1608--52.  The first of these systems is unique among the X-ray
binaries: It is the first known accretion-driven millisecond X-ray
pulsar (pulse period 2.5 ms; Wijnands \& van der Klis 1998).

\bigskip

\centerline{\bf Table 1:  BLACK HOLE X-RAY NOVAE}

\medskip

\begin{center}
\begin{tabular}{|c||c|c|l|} \hline
System&f(M/M$_{\odot}$)=PK$^{3}/2\pi$G&M$_{1}$ (est.)&Reference \\ 
\hline\hline
XTE J1859+226&$7.4\pm 1.1$&$10\pm 3$&1 \\ \hline
XTE J1550--564&$6.86\pm 0.71$&$>7.4$&2 \\ \hline
GS 2023+338&$6.08\pm 0.06$&$12\pm 2$&3 \\ \hline
XTE J1118+480&$6.0\pm 0.3$&$7\pm 1$&4,5 \\ \hline
GS 2000+250&$4.97\pm 0.10$&$10\pm 4$&3 \\ \hline
H1705--250&$4.86\pm 0.13$&$4.9\pm 1.3$&3 \\ \hline
GRS 1009--45&$3.17\pm 0.12$&$4.2\pm 0.6$&6 \\ \hline
GRS 1124--683&$3.01\pm 0.15$&$7\pm 3$&3 \\ \hline
A0620--00&$2.91\pm 0.08$&$10\pm 5$&3 \\ \hline
SAX J1819.3--2525&$2.74\pm 0.12$&$10.2\pm 1.5$&7 \\ \hline
GRO J1655--40&$2.73\pm 0.09$&$7\pm 1$&8 \\ \hline
GRO J0422+32&$1.21\pm 0.06$&$10\pm 5$&3 \\ \hline
4U 1543--47&$0.22\pm 0.02$&$5\pm 2.5$ &3 \\ \hline\hline
\end{tabular}
\end{center}

\noindent \hspace*{0.4in}References:  (1) Filippenko \& Chornock 2001;
(2) Orosz et al. 2001a; \\
\hspace*{0.4in}(3) McClintock 1998; (4) McClintock et al. 2001a; \\
\hspace*{0.4in}(5) Wagner et al. 2001; (6) Filippenko et al. 1999; \\
\hspace*{0.4in}(7) Orosz et al. 2001b; (8) Shahbaz et al. 1999.

\bigskip

\section{Quiescent Luminosities of X-ray Novae}

In order to search for differences between X-ray novae containing
black hole candidates and those containing neutron stars, we have
studied X-ray observations of these systems in quiescence.  We have
concentrated on the quiescent phase because, based on the
advection-dominated accretion flow (ADAF) model described in \S4, we
expect the event horizon to be most clearly revealed in quiescence.

During outburst, X-ray novae often have luminosities approaching the
Eddington limit, which means that BHXN, because of their larger
masses, are typically a factor of $\sim 5$ brighter than NSXN.  During
quiescence, when advection is most likely to dominate the accretion
flow, the ADAF model predicts that the luminosities should differ by a
large factor in the opposite sense, i.e. the black hole systems should
be significantly dimmer than the neutron star systems, provided the
black holes have event horizons and the neutron stars have solid
surfaces.  It is this signature that we have looked for in the
observations.

The ADAF model predicts differences also in the X-ray spectrum between
BHXN and NSXN, but these may not be as easily discerned as the
differences in the luminosity.  A full study of X-ray spectra measured
with the {\it Chandra} X-ray Observatory (Weisskopf \& O'Dell 1997)
will be presented in Kong et al. (2001).

\bigskip

\vbox{
{\small
\begin{center}
Table~2:~Quiescent Luminosities of NSXN and BHXN \\
\medskip
\begin{tabular}{lcc} \hline \hline
\\
System & $P_{\rm orb}$ (hr) & $\log [L_{\rm min}]~({\rm erg~s^{-1}})$ 
\\ \\ (1)
& (2) & (3) \\
\\ \hline \\
$\circ$ SAX J1808--365 & $2.0^a$  & 	31.5$^{b,c}$ \\
$\circ$ EXO 0748--676    & 3.82           & 34.1    \\
$\circ$ {\bf 4U 2129+47}      & 5.2       & {\bf 33.0$^d$},32.8    \\
$\circ$ MXB 1659--298         & 7.1       &  $<32.3^e$    \\
$\circ$ H1608--52           	&12$^f$   & 33.3   \\
$\circ$ {\bf Cen X--4}         & 15.1      & {\bf 32.3$^g$},32.4   \\
$\circ$ {\bf Aql X--1}         & 19       & {\bf 33.6$^h$},33.2$^i$   \\
\\
\hline
\\
$\bullet$ {\bf GRO J0422+32}    & 5.1   &{\bf  30.9},$<31.6$   \\
$\bullet$ {\bf A0620--00}       & 7.8   &{\bf 30.5},30.8$^j$   \\
$\bullet$ {\bf GS~2000+25 }     & 8.3   &{\bf 30.4},$<32.2$    \\
$\bullet$ GS1124--683            & 10.4  &$< 32.4$     \\
$\bullet$ H1705--250             & 12.5  &$< 33.0$     \\
$\bullet$ {\bf 4U 1543--47}             & 27.0  &$<${\bf 31.5},$<33.3$  \\
$\bullet$ {\bf GRO J1655--40}           & 62.9  &{\bf 31.3},32.4   \\
$\bullet$ {\bf V404 Cyg}        & 155.3 &{\bf 33.7},33.1$^k$    \\
\\
\hline

\end{tabular}
\end{center}
NOTES --- {\it Chandra\/} Measurements are in {\bf bold face}. 
(1) $\circ$ indicates a neutron star primary and $\bullet$ a black
 hole primary.  
(2) Orbital period. 
(3) Luminosity in quiescence in the 0.5--10 keV band.
\noindent 
$^a$Chakrabarty \& Morgan 1998; 
$^b$Wijnands et al. 2001;
$^c$Dotani, Asai, and Wijnands 2000; 
$^d$Nowak et al. 2001;
$^e$Wijnands 2001, this source has been added to this table for the
first time here; 
$^f$Wachter 2000 reports a new 12~hr period, intermediate between the
previously reported 98.4~hr (Ritter \& Kolb 1998) and 5~hr (Chen et
al. 1998) periods; 
$^g$Rutledge et al. 2001a;
$^h$Rutledge et al. 2001b; 
$^i$Recomputed for $d=5$~kpc from Narayan et al. (1997);
$^j$Recomputed for $\Gamma = 2$ from Narayan et al. (1997); 
$^k$Kong 2000.
\hfill\\
}
}

Table~2 presents all the available measurements of quiescent
luminosities of BHXN and NSXN with known orbital periods.  This table
is an updated version of that in Garcia et al. (2001, see also
Narayan, Garcia \& McClintock 1997, Asai et al. 1998, and Menou et
al. 1999, hereafter M99), and includes all sensitive measurements
published in the literature and recent measurements made with {\it
Chandra} (the latter are in bold face).  We note that {\it Chandra}
has been particularly important for measuring the quiescent
luminosities of the extremely dim black hole systems.  The exquisitely
polished X-ray mirror (van Speybroeck et al. 1997) at the heart of the
observatory allows a $\sim 100$ fold increase in sensitivity over
previous X-ray observatories, permitting us finally to observe the
quiescent state of BHXN.

While there are now 13 dynamically confirmed BHXN (Table~1) and a
number of NSXN with known orbital periods, sensitive observations of
the quiescent X-ray luminosity exist for only the subset of 8 BHXN and
7 NSXN listed in Table~2.  We do not include measurements made with
non-imaging detectors, or those of very short duration with imaging
detectors, as the resulting upper limits are so far above the
detections that they merely cloud the comparison (see Chen et al. 1998
and M99).

We note that in order to compute luminosities, we must assume an X-ray
spectral shape and must estimate the hydrogen column density to the
source to correct for the low energy absorption of the observed flux.
If the spectrum is not measured we assume a power-law spectrum with a
photon index $\Gamma = 2$, which is consistent with the spectral
results obtained for the three brightest sources (Kong et al. 2001).
We estimate the column density from the optically determined
interstellar absorption.  The computed luminosities are not very
dependent on these assumptions.

Figure~1 displays the Eddington-scaled quiescent luminosities of NSXN
and BHXN as a function of the binary orbital period $P_{orb}$.  We
scale the luminosity by the Eddington luminosity $L_{\rm Edd}$ because
this is the maximum luminosity of an accreting mass and is the natural
scale to measure the accretion luminosity.  If the luminosity is
greater than $L_{\rm Edd}$, radiation pressure acting on the accreting
gas will overwhelm the inward pull of gravity and will prevent the gas
from accreting.  For ionized hydrogen with opacity dominated by
Thomson scattering, $L_{\rm Edd}=1.25\times10^{38} (M_1/M_\odot) ~{\rm
erg\,s^{-1}}$.

Versions of Fig. 1 were presented in several earlier papers, e.g.
Narayan et al. (1997b), Garcia et al. (1998), M99, and in each case we
claimed that the data indicated that BHXN are dimmer than NSXN in
quiescence (but see Chen et al. 1998).  The version shown here, which
is taken from Garcia et al. (2001) and includes a number of
measurements with {\it Chandra}, significantly strengthens our earlier
claims.  We see that BHXN are dimmer than NSXN with comparable orbital
periods by a factor of 100 or more (see the non-hatched region of the
plot).  Such a large difference is not expected unless there is an
important qualitative difference in the nature of the accretors in the
two kinds of system, e.g. an event horizon versus a surface.  This is
discussed in \S4.

\begin{figure}[t]
\centerline{\psfig{figure=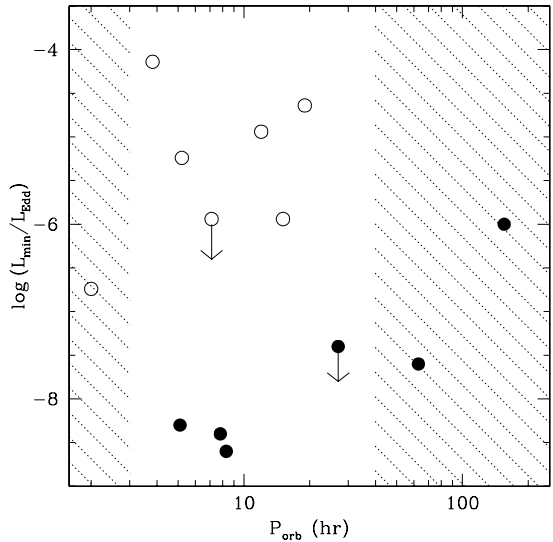,width=3.0in}}
\caption{Quiescent luminosities of BHXN (filled circles) and NSXN
(open circles) plotted against the binary orbital period.  Only the
lowest quiescent detections or upper limits are shown.  We have not
plotted the upper limits for GS1124--683 and H1705--250 because of the
relative insensitivity of the currently available observations of
these sources (neither has been observed with {\it Chandra}).  The
non-hatched region includes both BHXN and NSXN and allows a direct
comparison of BHXN and NSXN.  The hatched region on the left has no
BHXN and the region on the right has no NSXN.  These regions of the
plot are less useful.\label{fig:radish}}
\end{figure}


The use of $P_{orb}$ along the abscissa in Fig.~1 requires some
discussion.  At the short orbital periods characteristic of the X-ray
novae in our sample, angular momentum loss through gravitational
radiation is expected to be the dominant mechanism driving mass
transfer from the secondary.  As M99 showed, the Eddington-scaled mass
transfer rate is then determined primarily by $P_{orb}$ and is
insensitive to the mass $M_1$ of the compact star.  Therefore, the
Eddington-scaled mass accretion rate onto the compact star is likely
to be roughly similar for BHXN and NSXN of similar $P_{orb}$.  In the
non-hatched region of Fig.~1, we have several BHXN and NSXN with
similar orbital periods and therefore similar Eddington-scaled mass
accretion rates.  In the standard thin accretion disk model (\S4), we
would expect the Eddington-scaled luminosities of the systems to be
similar.  The data are clearly inconsistent with this expectation.  As
we discuss in \S4, the difference in luminosity is expected under the
ADAF model, but only if black holes have event horizons.

At the long orbital periods of two of the BHXN, V404~Cyg and
GRO~J1655--40 (the two rightmost filled circles in Fig.~1), nuclear
evolution rather than gravitational radiation is expected to drive the
mass transfer, and the mass transfer rate is no longer uniquely
determined by $P_{orb}$.  Further, there are no NSXN with $P_{orb}$ in
this range.  For both reasons these particular BHXN are less useful
for the purposes of our comparison.  One NSXN with a very short
orbital period, SAX J1808.4--3658 (the leftmost open circle in Fig.~1),
is similarly not useful since there is no comparison BHXN with a
comparable period.  This NSXN is also anomalous in being the only
known millisecond X-ray pulsar (see \S4.3).

\section{Advection-Dominated Accretion and the Event Horizon}

To understand and interpret observations of X-ray binaries it is
necessary to model the hydrodynamics and radiation processes of
viscous gas orbiting in the gravitational potential of a point mass.
The best-known accretion model is the thin accretion disk, which was
originally developed by Shakura \& Sunyaev (1973), Novikov \& Thorne
(1973) and Lynden-Bell \& Pringle (1974).  The orbiting gas in this
model is relatively cold (which is why it forms a geometrically thin
disk) and optically thick.  The gas radiates essentially all the heat
energy produced by viscous dissipation.  Therefore, for accretion onto
a neutron star or a black hole, the luminosity is of order $0.1\dot
Mc^2$ where $\dot M$ is the mass accretion rate.  Because the
accreting gas is optically thick, the radiation is blackbody-like
(with a radius-dependent temperature) and the spectrum consists of a
``multi-color blackbody.''  For an accretor with $M\sim 1-15M_\odot$
(as appropriate for a neutron star or black hole X-ray binary) which
is accreting at close to the Eddington rate, the effective temperature
of the blackbody emission is in the range $1-3$ keV, which is in the
soft X-ray band.

Most X-ray binaries with luminosities above a few percent of the
Eddington limit radiate primarily in soft X-rays and have spectra that
resemble the predictions of the thin disk model (Mitsuda et al. 1984;
Makishima et al. 1986).  Such sources are said to be in the ``high
soft'' spectral state.  The sources do have some hard X-ray emission,
which is thought to arise in a hot corona (Haardt \& Maraschi 1991,
1993; Haardt, Maraschi, \& Ghisellini 1994), but the bulk of the
emission is consistent with a multicolor blackbody from a standard
thin disk.

At luminosities below a few percent of Eddington, X-ray binaries
switch to a different state, the ``low hard'' state (e.g. Tanaka \&
Shibazaki 1996; Barret et al. 2000), in which the soft X-ray emission
decreases substantially and is replaced by a hard power-law component.
The switch is particularly dramatic in BHXN, but it is seen also in
other X-ray binaries, e.g. persistent black hole sources like Cyg X--1
(Cui et al. 1998) and GX339--4 (Kong et al. 2000), and NSXN such as Cen
X--4 (Bouchacourt et al. 1984) and Aql X--1 (Campana et al. 1998;
Verbunt et al. 1994).  The spectrum in the low state is most naturally
explained as due to Comptonization of soft photons by a hot
optically-thin plasma.  The density and temperature of the plasma, as
implied by the observations, are very different from those predicted
for a thin disk.  Thus, the low state must correspond to a different
mode of accretion than the thin disk.

\subsection{Advection-Dominated Accretion}

Apart from the thin accretion disk, a second stable accretion flow
solution is known, the so-called advection-dominated accretion flow
(ADAF).  The solution was discovered by Ichimaru (1977), and some
aspects of it were discussed by Rees et al. (1982).  However, apart
from these two studies, hardly any work was done between 1977 and
1994.  Modern interest in the ADAF model began with the papers of
Narayan \& Yi (1994, 1995a,b), Abramowicz et al. (1995) and Chen et
al. (1995).

The key feature of an ADAF (see Narayan, Mahadevan \& Quataert 1998
and Kato, Fukue \& Mineshige 1998 for reviews) is that the heat energy
released by viscous dissipation is not radiated immediately, as in a
thin disk, but is stored in the gas as thermal energy and advected
with the flow --- hence the name.  Because of the large amount of
energy involved, the gas becomes extremely hot, with ions (mostly
protons) reaching nearly the virial temperature, $T_i\sim10^{12}\,{\rm
K}/r$, where $r$ is the radius in Schwarzschild units.  At such
temperatures, the gas is very distended (the configuration is no
longer ``thin'') and the density is very low.

The low density of an ADAF is a key property, since it causes
particle-particle interactions to be infrequent.  This leads to the
radiative inefficiency that is characteristic of an ADAF.  In detailed
models published in the literature, the accreting gas is treated as a
two-temperature plasma, with the electrons being cooler than the ions,
though still quite hot: $T_e>10^9$ K.  Heat energy from viscous
dissipation is assumed to go primarily to the ions which do not have
any efficient channel to radiate.  Energy has to be transfered from
the ions to the electrons before it can be radiated, but the transfer
is inefficient at the low densities found in an ADAF.  The radiative
efficiency of the electrons too is poor at sufficiently low densities.
Both effects cause the gas to become advection-dominated.

The ADAF solution has been shown to be thermally stable (Narayan \& Yi
1995b; Kato et al. 1997; Wu 1997), an important result.  Prior to the
development of the ADAF model, another hot accretion solution was
discovered by Shapiro, Lightman \& Eardley (1976), but that solution
turned out to be violently thermally unstable (Piran 1978) and
therefore not viable.  The ADAF is presently the only stable hot
solution known.

Detailed calculations (Narayan \& Yi 1995b; Esin, McClintock \&
Narayan 1997) show that the optically-thin two-temperature ADAF
solution is allowed only for Eddington-scaled mass accretion rates
$\dot m\equiv\dot M/\dot M_{\rm Edd}<\dot m_{\rm crit}\sim 0.01-0.1$.

For completeness, we note that at very high $\dot m>1$, a second ADAF
branch occurs (Katz 1977; Begelman 1978; Begelman \& Meier 1982;
Eggum, Coroniti \& Katz 1988; Abramowicz et al. 1988) in which
radiation is trapped and radiation-pressure dominates.  That solution
branch is not of interest for this article.

\subsection{ADAFs and X-ray Binaries}

The low-$\dot m$, two-temperature ADAF model described in \S4.1 has
three properties which make it attractive for applications to X-ray
binaries: high electron temperature, low density, thermal stability.
The first two properties ensure that the spectrum will be dominated by
Compton scattering rather than blackbody emission.  Many X-ray
binaries have X-ray spectra that are dominated by Compton scattering,
e.g. sources in the low hard state mentioned earlier.  The ADAF model
finds a natural application to these sources.  The soft photons for
Comptonization are supplied either by synchrotron emission from the
ADAF itself (Narayan \& Yi 1995b) or by thermal radiation from an
external thin disk (e.g. Esin et al. 1997) or from the heated surface
of a neutron star (e.g. Narayan \& Yi 1995b).

As mentioned above, the ADAF solution is viable only for somewhat low
mass accretion rates: $\dot m<\dot m_{\rm crit}\sim0.01-0.1$.  The low
hard spectral state of X-ray binaries similarly occurs only for
relatively low luminosities: $L< 0.01-0.1L_{\rm Edd}$.  It is,
therefore, natural to assume that the low state corresponds to an
ADAF.  Spectra computed with the ADAF model match observations of
black hole binaries in the low state quite well (Esin et al. 1997,
1998), thus bolstering the association.

Narayan (1996) and Esin et al. (1997) proposed the following scenario
to account for the broad qualitative features of the various spectral
states of X-ray binaries.  For $\dot m> \dot m_{\rm crit}$ , they
postulated that the accretion occurs via a thin disk, possibly with a
corona.  The system is then in the high soft state.  However, once
$\dot m<\dot m_{\rm crit}$, they suggested that the inner regions of
the thin disk ``evaporate'' and are replaced by an ADAF zone.
(Technically, the thin disk solution is viable even for $\dot m<\dot
m_{\rm crit}$, but these authors assumed that once the ADAF solution
is allowed nature picks the ADAF in preference to the thin disk.)
With decreasing $\dot m$, the size of the ADAF zone increases and the
spectrum becomes progressively more and more dominated by the hard
Comptonized emission of the ADAF; this is the low hard state.  The key
feature of the model is the formation of a central hole in the thin
disk which is filled with hot ADAF gas.  This feature has been
confirmed spectacularly by recent observations of the BHXN XTE
J1118+480 (McClintock et al. 2001b; Esin et al. 2001).  Esin et
al. (2001) have shown that the hole in the thin disk in this source
must have a radius $\sgreat\ 55$ Schwarzschild radii, rather than the
3 Schwarzschild radii observed in the high soft state.

According to the above scenario, as the mass accretion rate decreases,
the thin disk recedes to larger and larger radii and its luminosity
falls rapidly.  The emission from the ADAF also decreases
substantially because the radiative efficiency decreases with
decreasing $\dot m$ (due to the lower density).  The luminosity of the
source thus falls steeply; roughly $L\sim \dot m^2$ (Narayan \& Yi
1995b).  At sufficiently low $\dot m$, the luminosity corresponds to
that observed in the quiescent state of X-ray novae, and in this state
the ADAF zone is expected to be quite large ($\sim10^3-10^4$
Schwarzschild radii in models).  According to this picture, the
quiescent state is not distinct from the low state, but is just an
extreme version of it.  We note that there is no observational
evidence for an abrupt spectral transition between the low state and
the quiescent state, such as is seen when a source goes from the high
state to the low state; the observations are thus consistent with the
basic picture.

In the interests of pedagogy, we have discussed first the low state
and introduced the quiescent state as an extreme example of the low
state.  However, historically, it was the quiescent state that was
first convincingly linked to the ADAF model.  Narayan, McClintock \&
Yi (1996; see also Lasota, Narayan \& Yi 1996) argued that the BHXN
A0620--00 in quiescence has a spectrum that is inconsistent with a thin
disk model, and showed that the spectrum can be explained with a model
that has an ADAF at small radii and a thin disk at large radii.
Narayan, Barret \& McClintock (1997) modeled the spectrum of the BHXN
V404 Cyg (which had better X-ray data in quiescence) and showed that
the ADAF model again fits the observations quite well.  Hameury et
al. (1997, see also Menou et al. 2000) showed that the X-ray delay
observed by Orosz et al. (1997) during the onset of an outburst of the
BHXN GRO J1655--40 is best understood by invoking a hole in the thin
disk, as postulated in the ADAF model.  Similar X-ray delays have been
seen also in outbursts of the NSXN Aql X--1 (Jain et al. 2001a; Shahbaz
et al. 1998) and the BHXN XTE J1550--564 (Jain et al. 2001b), suggesting
that these sources too, when in quiescence, have a central ADAF zone.

\subsection{The Event Horizon}

What is the relevance of all this to the event horizon?  The
connection is simple.  Recall that the key feature of an ADAF is that
most of the energy released by viscous heating ($\sim0.1 \dot Mc^2$)
is retained in the gas and advected to the center; only a small
fraction of the energy is radiated.  Indeed, as already noted, the
fraction that is radiated decreases with decreasing $\dot m$.  For
values of $\dot m$ appropriate to the quiescent state of X-ray novae,
the fraction that is radiated is likely to be extremely small (less
than $10^{-2}$ and as low as $10^{-4}$ in some models, compared to
$\sim 1$ for standard thin disk accretion).  What happens to the
$\sim0.1\dot Mc^2$ of advected energy?

If the accretor is a black hole, the advected energy disappears
through the horizon as the gas falls in.  But if the accretor is a
neutron star, the accreted gas will come to rest on the stellar
surface, and since the density at this point will be many orders of
magnitude higher than in the ADAF the gas will radiate its stored
energy.  Thus, for accretion with low $\dot m$ via an ADAF, the
luminosity is much higher when the central accretor has a surface than
when the accretor has an event horizon.

This simple consequence of the ADAF model was pointed out by Narayan
\& Yi (1995b) and first tested by Narayan et al. (1997b) who compared
the quiescent luminosities of BHXN and NSXN.  Section 3 discusses the
current status of the test (see Fig.~1).  The data show very clearly
that quiescent BHXN are {\it significantly} dimmer than quiescent
NSXN.  Since we expect the two kinds of sources to have similar
Eddington-scaled accretion rates $\dot m$ (\S3), we interpret the
observations as the first clear evidence for the presence of event
horizons in black hole candidates.  Note that we need both an ADAF and
an event horizon to explain the unusually low luminosity of the BHXN;
neither one alone is enough.

While Fig. 1 is qualitatively consistent with the predictions of the
ADAF model, the situation is not so satisfactory on a quantitative
level.  The basic ADAF model predicts that the difference in
luminosity between BHXN and NSXN should be much larger than the factor
of $\sim 100$ seen in the data.  M99 discussed this discrepancy and
suggested that perhaps NSXN are anomalously dim because of propeller
action of a spinning magnetized neutron star.  Robertson \& Leiter
(2001) describe a model in which the transition from the high state to
the low state is the result of propeller action.  They model X-ray
novae in quiescence as propeller systems with significant
magnetospheric emission.

There is good evidence for rapid spin in a few dozen X-ray pulsars
(White and Zhang 1997), but most of these are not NSXN.  Only in one
NSXN, namely SAX J1808.4--3658 (Wijnands \& van der Klis 1998), is
there direct evidence for both rapid spin and a strong enough magnetic
field to influence the accretion flow.  It is interesting that this
source has the lowest luminosity among all the quiescent NSXN shown in
Fig.~1 (the leftmost open circle); perhaps this source is anomalously
dim due to a propeller.  There is indirect evidence for a strong
magnetic field in Aql X--1 (Campana et al. 1998), and the source
exhibited nearly coherent oscillations during a Type I burst with a
frequency of 549 Hz (Zhang et al. 1998).  However, Chandler \&
Rutledge (2000) question the presence of a propeller in this source
during quiescence.  The propeller explanation appears to be even less
plausible for the other five NSXN shown in Fig.~1 (but see Robertson
\& Leiter 2001).

Campana \& Stella (2000) argue that even though a propeller may
prevent gas from reaching the neutron star, the gas must still accrete
down to the magnetospheric radius before being expelled and should
radiate the binding energy corresponding to this radius.  The source
should thus be quite bright.  This argument is valid if the accreting
gas is radiatively efficient.  However, the argument fails if the
accretion occurs via an ADAF and if the gas is also ejected by the
propeller in an advection-dominated state.

Another approach to understanding Fig.~1 quantitatively is based on
recent developments in ADAF theory, as we now describe.

\subsection{Recent Technical Developments}

Narayan \& Yi (1994, 1995a) noted the curious fact that the Bernoulli
parameter, namely the sum of the potential energy, kinetic energy and
enthalpy, of the accreting gas in an ADAF tends to be positive.  These
authors recognized that this means the gas is technically not bound to
the accreting mass, and they suggested that an ADAF may have strong
outflows.  Blandford \& Begelman (1999) developed the idea further and
argued that the mass accretion rate in an ADAF may decrease rapidly
with radius, so that only a small fraction of the available mass
reaches the central object.  Numerical simulations by Igumenshchev and
collaborators (Igumenshchev, Chen \& Abramowicz 1996; Igumenshchev \&
Abramowicz 1999, 2000) indicate strong outflows whenever the viscosity
in the gas is very large; however, the required viscosity coefficient
appears to be uncomfortably large.

Preliminary MHD simulations of ADAFs seem to indicate strong outflows
(Stone \& Pringle 2001; Hawley, Balbus \& Stone 2001), but in our
opinion the interpretation of the results is unclear.  The flows could
be viewed either as ADAFs with outflows or as the MHD version of the
convective flows discussed below (see Machida, Matsumoto \& Mineshige
2001).

Another interesting fact is that ADAFs are convectively unstable
(Narayan \& Yi 1994, 1995a).  Viscous dissipation in an ADAF, coupled
with the low radiative efficiency, causes the entropy of the gas to
increase inward.  The negative radial entropy gradient makes the gas
subject to a convective instability.  (Rotation also plays a role in
the instability criterion, as discussed in the papers cited below.)
Numerical simulations (Stone, Pringle \& Begelman 1999; Igumenshchev
\& Abramowicz 1999, 2000; Narayan, Igumenshchev \& Abramowicz 2000;
Igumenshchev, Abramowicz \& Narayan 2000) have confirmed the
convective instability in hydrodynamic flows with relatively low, and
``reasonable,'' values of the viscosity coefficient.  ADAFs with
convection are called convection-dominated accretion flows, or CDAFs
(Quataert \& Gruzinov 2000; Narayan et al. 2000).

The most interesting result to come out of the numerical work is that
the radial structure of a CDAF (especially the density and radial
velocity) is very different from that of a non-convective ADAF.  The
key physical reason for this difference was identified by Narayan et
al. (2000) and Quataert \& Gruzinov (2000): Convective turbulence does
not behave like ordinary viscosity, which moves angular momentum
outward; rather, convective turbulence moves angular momentum inward
(this concept was discussed earlier for thin disks by Ryu \& Goodman
1992 and Stone \& Balbus 1996, and for rotating stars by Kumar,
Narayan \& Loeb 1995).

An important result of the recent work on CDAFs is the following: For
the same outer boundary conditions, the mass accretion rate $\dot m$
onto the central accretor is substantially less in a CDAF than in a
standard ADAF.  Loeb, Narayan \& Raymond (2001) have shown that data
on quiescent cataclysmic variables (CVs: accreting white dwarfs) and
NSXN are consistent with these objects having CDAFs rather than ADAFs.
There is, therefore, some justification for applying the CDAF model to
quiescent X-ray novae.

Balbus (2001) has argued recently that convection in MHD is very
different from pure hydrodynamic convection, and has questioned the
validity of the assumptions behind the above-cited work on CDAFs.
This needs further study.

Whether accretion proceeds via an ADAF or a hydrodynamic CDAF or an
MHD analog of these models or any other radiatively inefficient flow,
whatever mass reaches the center will do so with nearly virial
temperature and thus a great deal of thermal energy.  Consequently, by
the argument given earlier, a black hole system will be significantly
dimmer than a neutron star system.  Thus, the interpretation of Fig.~1
as evidence for the event horizon is qualitatively valid for any
radiatively inefficient flow.  But the quantitative predictions will
differ from one model to another.  Since the radiative efficiency of a
CDAF is larger than that of an ADAF with the same mass accretion rate,
the luminosity difference between BHXN and NSXN will be smaller in the
case of a CDAF than for an ADAF.  Indeed, Abramowicz \& Igumenshchev
(2001) have argued (following the work of Ball, Narayan \& Quataert
2001) that the observed factor of 100 difference between BHXN and NSXN
in Fig.~1 is consistent with the CDAF model.

As discussed earlier, applications of the ADAF model to X-ray binaries
make use of a flow geometry consisting of two zones separated at a
transition radius $r_{tr}$ (Narayan et al. 1996; Narayan 1996; Esin et
al. 1997).  For $r<r_{tr}$, there is a hot two-temperature zone which
is either an ADAF or a CDAF.  For $r>r_{tr}$, the accretion occurs
partially as a thin disk and partially as a hot corona (the corona
being modeled as an ADAF-like hot zone).  A similar geometry was
proposed many years earlier by Shapiro et al. (1976; see also Wandel
\& Liang 1991), except that they used a different accretion solution
for the hot zone.  Their solution turned out to be unstable (Piran
1978), but the basic idea is the same.

Why should the inflowing gas switch from a thin disk to a hot flow at
$r=r_{tr}$?  Meyer \& Meyer-Hofmeister (1994) proposed for CVs a
mechanism in which the disk is heated by electron conduction from the
corona; the heat flux causes the cold disk to evaporate into the hot
phase.  A number of studies have investigated this and related
mechanisms for X-ray novae (Honma 1996; Dullemond \& Turolla 1998; Liu
et al. 1999; Rozanska \& Czerny 2000; Czerny, Rozanska \& Zycki 2000;
Meyer, Liu \& Meyer-Hofmeister 2000; Spruit \& Deufel 2001) and have
shown that the idea works fairly well.  However, the models are not
yet able to make reliable quantitative predictions of the dependence
of $r_{tr}$ on $\dot m$.

One final comment concerns particle heating in advection-dominated
flows.  Many ADAF models published in the literature assume that most
of the viscous heat energy goes into the ions and that the electrons
receive only a small fraction (say 1\%) of the energy (e.g. Narayan et
al. 1997a; Esin et al. 1997).  This assumption has been questioned by,
among others, Bisnovatyi-Kogan \& Lovelace (1997; see also Phinney
1981).  Detailed calculations of particle heating by turbulent eddies
in MHD flows (Quataert 1998; Gruzinov 1998; Blackman 1999; Quataert \&
Gruzinov 1999) show that the assumed low level of electron heating is
plausible if the magnetic field is weak (plasma $\beta>10$), but is
unlikely if the field has near-equipartition strength.  Recent work
has shown that, if accretion proceeds with a strong outflow or via a
CDAF, then one can obtain reasonable spectra that match observations
even if as much as half the viscous heat energy goes into the
electrons (Quataert \& Narayan 1999; Ball et al. 2001).  This is
welcome since it implies that the CDAF model does not require special
assumptions on particle heating (as the original ADAF model did).
However, detailed modeling of individual sources with the CDAF model
has not yet been attempted.

To summarize, recent developments in ADAF theory have introduced new
ideas, principally the CDAF and role of MHD, and have identified
promising directions for further study.  These developments may
provide a better quantitative match between model predictions and
observations of the luminosity.  Model predictions of spectra are
awaited.

\section{Other Explanations of the Data on Quiescent X-ray Novae}

While we have concentrated on interpreting the X-ray data in the
context of the ADAF model, several alternate interpretations of the
difference in the X-ray luminosities of quiescent BHXN and NSXN have
been put forward.  We comment on the strengths and weakness of these
alternate explanations.

\subsection{Optical/UV Luminosity}

Campana and Stella (2000) argue that we have incorrectly assumed that
the X-ray luminosity is an accurate measure of the accretion rate near
the black hole or neutron star.  Motivated partially by some early
ADAF models, they suggest that the optical and UV luminosity also
originates near the central object, and that this luminosity should be
included in the comparison. The non-stellar optical/UV luminosity is
much greater than the X-ray luminosity.  When it is included, the
difference in luminosity between BHXN and NSXN largely disappears.

However, there are some problems with this argument.  The level of UV
and optical emission generated by an ADAF depends on the details of
the flow.  If there are strong winds, or if the flow occurs as a CDAF,
the optical/UV emission is greatly suppressed (Quataert \& Narayan
1999; Ball et al. 2001).  A comparison of the luminosities of
quiescent accreting neutron stars and white dwarfs indicates that
these systems do have winds or behave like CDAFs (Loeb et al. 2001).
Thus, the observed optical and UV emission is unlikely to be generated
in the hot gas close to the central accretor.

The point where the mass transfer stream from the companion star
impacts the outer edge of the accretion disk (the ``hot spot'')
typically generates a large fraction of the UV and optical emission in
CVs, and we suggest that it does so in quiescent X-ray novae as well.
An 0.25 mag modulation on an orbital time scale has been observed in
the far-UV flux of the BHXN A0620--00 (McClintock 2000), indicating
that the far-UV flux in this source may originate in the hot spot.  If
the optical/UV luminosity originates in the outer accretion disk
and/or the hot spot, then the fact that the optical/UV luminosities of
BHXN and NSXN are similar provides observational confirmation that the
mass transfer rates (at the outer edge of the disk) in the two kinds
of system are similar, which is an important assumption behind our
interpretation of Fig.~1.  We conclude that further observations are
needed to determine the origin of the optical/UV emission in quiescent
X-ray novae.

\subsection{Models That Do Not Involve Accretion}

Fundamental to our interpretation of Fig.~1 as evidence for event
horizons is the assumption that the quiescent luminosity of X-ray
novae is powered by accretion.  Could this assumption be wrong?

\subsubsection{i. Coronal Emission from Black Hole Secondaries}

\begin{figure}
\centerline{\psfig{figure=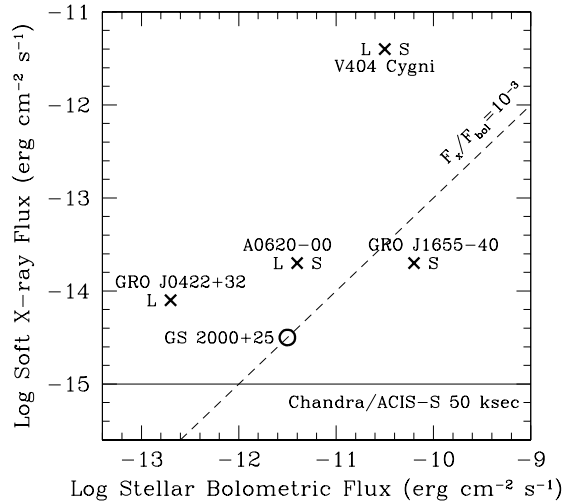,width=3.5in}}
\caption{After Bildsten \& Rutledge (2000).  The solid line indicates
the limiting flux for a 50 ksec observation with the Chandra/ACIS-S,
and the dashed line indicates the limiting soft X-ray flux of the
coronal model.  Systems above the dashed line have X-ray luminosities
too high to be produced by a stellar corona.  The Chandra observations
are indicated by the open circle and the four crosses.  The open
circle for GS~2000+25 indicates that this system has a quiescent
luminosity consistent with the coronal model.  The crosses for the
other four systems, GRO~J0422+32, A0620--00, V404~Cygni and
GRO~J1655--40, indicate that these sources are not consistent with the
coronal model.  An ``L'' indicates that the luminosity is too high and
an ``S'' indicates that the spectrum is inconsistent.  The system 4U
1543--47 has not been plotted because it has a radiative secondary
which is not expected to have coronal emission.
\label{fig:BR}}
\end{figure}

\vspace*{-2pt}

Bildsten \& Rutledge (2000) suggest that much of the X-rays observed
from quiescent BHXN may be produced by a rotationally enhanced stellar
corona in the secondary star, as seen in tidally locked binaries such
as the RS CVn systems.  This suggestion has been criticized by Lasota
(2000).  In addition, the luminosity and spectral evidence available
for four of the six BHXN observed by Chandra rule strongly against the
proposal.

Figure 2 compares the luminosities of quiescent BHXN with the
predictions of the coronal model.  The quiescent luminosity of GRO
J0422+32 exceeds the maximum prediction of the coronal model by a
factor of $\sim 25$, and V404~Cyg exceeds the limit by a factor of
$\sim 60$.  A0620--00 is a factor of $\sim 6$ above the coronal
prediction, which is still a significant discrepancy since the
prediction corresponds to the maximum likely level of coronal
emission.

Turning to the spectral evidence, we find (Kong et al. 2001) that the
X-ray spectra of V404 Cyg, GRO J1655--40 and A0620--00 are harder
(equivalently hotter) than typical spectra of stellar coronae (e.g.,
Dempsey et al.  1993).  Thus, in the three systems for which the data
are of sufficient quality to allow us to measure the X-ray spectrum,
we can show that it is unlikely that coronal emission dominates the
observed X-rays.

It is worth noting that the four systems discussed above, for which
coronal emission is ruled out, cover the full range of orbital period
and stellar bolometric flux.  It therefore seems unlikely that there
is some particular region of parameter space where the coronal model
applies.  In comparison, the ADAF model is consistent with all the
observations, covering the full parameter space (Narayan et al. 1996,
1997a; Lasota 2000).

In the case of GS 2000+25, the luminosity is consistent with the
coronal hypothesis, but the source is too faint to obtain useful
spectral information.  The final BHXN observed by Chandra, 4U
1543--47, has an A2V radiative secondary (Orosz et al. 1998), and is
not expected to have a corona.  For this reason, it is not relevant
for this discussion and is not included in Fig.~2.  Thus, of the five
potentially corona-dominated BHXN observed by Chandra, only one (GS
2000+25) is consistent with the hypothesis that coronal emission
dominates the observed X-rays.

An obvious point to note is the following.  Emission from a stellar
corona will contribute at some level to the X-ray emission from
quiescent BHXN, as discussed by Bildsten \& Rutledge (2000).  If in a
few cases this level is significant, then the accretion luminosities
of the black holes must be even lower than our estimates and the
argument for event horizons would be further strengthened.

\subsubsection{ii. Incandescent Neutron Stars}

Brown, Bildsten \& Rutledge (1998) have suggested that the quiescent
luminosity of NSXN could be due to deep crustal heating of the neutron
star during outburst followed by cooling in quiescence.  The observed
quiescent luminosities of several NSXN are consistent with the
predictions of this model; also, the observed spectra reveal a soft
component that is well fit as thermal emission from a light element
atmosphere on a cooling neutron star (e.g. Rutledge et al. 1999,
2001).

Nevertheless, despite these successes, it appears that incandescence
alone cannot explain all the observations.  The rapid variability of
the prototypical NSXN Cen X--4 (Campana et al. 1997) is hard to
understand in a cooling model, and shows that at most $\sim 1/3$ of
the quiescent luminosity of this source is due to crustal cooling
(M99).  In addition, Cen X--4 and Aql X--1 have substantial power-law
tails in their spectra, carrying about half the total luminosity (Asai
et al. 1996; Campana et al. 1998).  This spectral component is
unlikely to arise from the surface of a cooling neutron star.

Both the variability and the power-law emission are expected in an
accretion model; indeed, even the thermal component in the spectrum is
natural in an accretion model --- it would correspond to that part of
the accretion luminosity radiated from the stellar surface.  It thus
seems likely that accretion accounts for a substantial fraction
(perhaps even most) of the quiescent X-ray luminosity in many NSXN.
Moreover, we should note that even if the fraction due to accretion is
modest, say half or a third of the total luminosity, the argument for
the event horizon is hardly affected since the luminosities of
quiescent BHXN and NSXN differ by a much larger factor $\sim100$.  In
this sense, the event horizon argument we have presented is robust,
since it is based on a very large signal that is insensitive to minor
effects at the level of a factor of 2.  Incidentally, the optical
variability of NSXN in quiescence (McClintock \& Remillard 2000; Jain
et al. 2000a; Ilovaisky \& Chevalier 2000) provides ample evidence that
mass transfer from the companion star continues during quiescence, so
there is no dearth of gas to feed the accretion.

As of this writing, no spectral models have been developed for
accreting neutron stars with ADAFs or CDAFs.  This is an area for
future research.

\subsubsection{iii. Pulsar Wind/Shock Emission}

Campana \& Stella (2000) have suggested that NSXN switch to a radio
pulsar-like mode in quiescence.  In this case, the X-ray luminosity is
expected to be of order the ``pulsar shock'' luminosity $\sim 2 \times
10^{32} ~{\rm erg\,s^{-1}}$, which is close to the observed level.
The observed power-law and thermal emission from quiescent NSXN is
naturally explained by this model: the shock produces the power law
emission and the neutron star surface radiates the thermal component.
The lack of any sign of periodicity in the quiescent emission (in any
electromagnetic band) could be a problem for this model.  SAX
J1808.4--3658 does show periodicity in outburst, but that emission is
clearly the result of accretion, not a pulsar wind/shock.

\subsection{Coronal X-ray Emission From the Outer Accretion Disk}

Recently, Nayakshin \& Svensson (2001) suggested that the X-ray
emission from quiescent X-ray novae arises in the outer accretion
disk, far from the compact object.  In their model, the mass accretion
rate inside of the radius $R_U$ where hydrogen first becomes partially
ionized falls as $\sim R^3$.  Thus the disk in the vicinity of the
compact object is almost devoid of mass and is dim at all wavelengths.
At larger radii, on the other hand, the accretion rates are much
larger, and the authors assert that the observed X-ray luminosities
can be easily reproduced assuming that the corona reprocesses $\sim
10$\% of the total accretion energy into X-rays.  The bulk of the
X-ray luminosity is thus emitted at $R_U$.  Given the observed optical
luminosity of the BHXN A0620--00 in quiescence $\sim 10^{32} ~{\rm
erg\,s^{-1}}$ (McClintock, Horne \& Remillard 1995), the model is able
to explain the observed X-ray luminosity of $\sim 10^{31} ~{\rm
erg\,s^{-1}}$.  Another plus for the model is the observed optical and
UV emission lines, which most likely originate in a chromospheric
layer between the cold disk and the hot corona.  The lines have a
luminosity $\sim 7 \times 10^{30}~{\rm erg\,s^{-1}}$ (McClintock \&
Remillard 2000), confirming that the chromosphere has an energy budget
of the same order as the observed X-ray luminosity.

While the Nayakshin \& Svensson (2001) model is similar to the ADAF
model in that the ultimate energy source is accretion, it predicts
that the X-ray emission originates at much larger radii --- in
A0620--00, the radius is $R_U \sim 5 \times 10^{10}$~cm.  Observations
of an eclipsing BHXN in quiescence could test this prediction.

Unfortunately an eclipsing BHXN has yet to be found.  However, the
model should equally well apply to quiescent CVs and there are
observations of eclipsing quiescent CVs.  In the three cases we know
of (HT Cas, Mukai et al. 1997; Z Cha, van Teeseling 1997; OY Car,
Pratt et al. 1999), the size of the X-ray emitting region is
comparable to the size of the accreting white dwarf.  The measurements
with the highest signal-to-noise ratio are those of HT Cas, for which
the emitting region is $< 8 \times 10^8$cm.  This limit is nearly 2
orders of magnitude smaller than $R_U$, and therefore presents a
problem for the model.

While Nayakshin \& Svensson (2001) present an alternate explanation
for the origin of the X-rays in quiescent BHXN, they do not attempt to
explain the $\sim 100$ times larger luminosity of quiescent NSXN.

Related to the model of Nayakshin \& Svensson (2001) is the one
proposed by Robertson \& Leiter (2001), in which the X-ray emission in
both NSXN and BHXN comes from a magnetospheric propeller zone away
from the compact star.  This model faces the same problem as the model
of Campana \& Stella (2000, \S5.2.iii): why do the observed quiescent
X-rays not show any pulsations?

\section{Discussion}

The advection-dominated accretion flow (ADAF) model provides a natural
framework to understand observations of X-ray novae at low
luminosities.  For luminosities below a few percent of the Eddington
luminosity, observations indicate that accretion flows in X-ray novae
are dominated by hot optically thin gas.  The only stable and
consistent model known for such hot flows is the ADAF (and the related
CDAF), and this model fits the observations quite well.

The ADAF model predicts that at very low mass accretion rates, such as
are found in quiescent BHXN and NSXN, there should be a large
difference in luminosity between accretors with an event horizon and
those with a surface (Narayan \& Yi 1995b).  A few years after this
prediction was made, a preliminary confirmation of the effect was
published (Narayan et al. 1997b) based on a modest amount of data.
The data today are far more convincing (Fig.~1) and indicate
unambiguously that quiescent BHXN are indeed very much dimmer than
quiescent NSXN.  We submit that this constitutes strong evidence for
the presence of event horizons in BHXN.

Section 5 describes several alternate explanations of the data.  Many
of these models have difficulties with various observations.  In
addition, most of the arguments are focused on disproving the ADAF
model rather than on disproving the presence of the event horizon.
Even if we accept these explanations, we still need to understand why
quiescent BHXN are a 100 times dimmer than quiescent NSXN.  If NSXN
are bright because of ``incandescent emission'' (\S5.2.ii), why are
BHXN not equivalently bright?  One reasonable explanation is that
black hole candidates have event horizons!  If the X-rays observed
from BHXN are due to coronal emission at some large radius in the disk
(\S5.3), why are NSXN so much brighter?  The originators of the disk
corona idea give no explanation.  If X-ray emission from NSXN is due
to a pulsar-like wind and a shock (\S5.2.iii), why do BHXN not have
similar winds and shocks?  The reason must be that black hole
candidates do not have the sort of magnetic fields that are invoked by
the authors for NSXN.  This is of course consistent with BHXN having
event horizons, since the ``no-hair theorem'' implies that a black
hole cannot have a frozen-in magnetic field.

As a final comment, we note that some of the best-studied
astrophysical phenomena that are thought to be associated with black
holes involve extremely luminous objects: X-ray novae in outburst,
active galactic nuclei, gamma-ray bursts.  All of these bright
phenomena are believed to be the result of matter falling into the
relativistic gravitational potential well of black holes or other
relativistic objects.  But none of the phenomena require (as far as
one can tell) an event horizon.  If this is all there is to a black
hole, one might as well call them ``bright holes''!

Physics tells us that what makes a black hole unique is the
``blackness'' of its event horizon.  In quiescent X-ray novae, for the
first time, we see evidence that candidate black holes are indeed
unusually black.  It would be disappointing if this long-sought
signature is not somehow connected with the event horizon.

The authors thank L. Bildsten, S.L. Robertson and R.E. Rutledge for
comments on the manuscript.  RN was supported in part by NASA grant
NAG5-10780.  MRG was supported by Contract NAS8-39073 to the Chandra
X-ray Center.  JEM was supported in part by NASA grant GO0-1105A.

\end{document}